\documentclass [a4paper]{article}
\title{Solving the measurement problem: de~Broglie-Bohm loses out to Everett}
\author{Harvey R Brown\thanks{Faculty of Philosophy, University of Oxford, 10 Merton Street, Oxford OX1 4JJ, U.K.; {\em harvey.brown@philosophy.ox.ac.uk.}}   \and David Wallace\thanks{Magdalen College, High Street, Oxford OX1 4AU, U.K.; {\em david.wallace@magdalen.ox.ac.uk.}}}
\begin{document}
\maketitle
\begin{abstract}
The quantum theory of de Broglie and Bohm  solves the measurement problem, but the hypothetical corpuscles play no role in the argument. The solution finds a more natural home in the Everett interpretation. 
\end{abstract}\

\bigskip

\emph{``If the quantum theory is to be able to provide a complete description of everything that can happen in the world $\dots$ it should also be able to describe the process of observation itself in terms of the wave functions of the observing apparatus and those of the system under observation. Furthermore, in principle, it ought to be able to describe the human investigator as he looks at the observing apparatus and learns what the results of the experiment are, this time in terms of the wave functions of the various atoms that make up the investigator, as well as those of the observing apparatus and the system under observation. In other words, the quantum theory could not be regarded as a complete logical system unless it contained within it a prescription in principle for how all these problems were to be dealt with}.'' D. Bohm~\cite{bohm51}, p. 583.

\section{Introduction} A common attitude amongst those that take the de Broglie-Bohm interpretation of quantum mechanics\footnote{On occasions for brevity we will refer to the theory simply as the Bohm theory. (For useful surveys of the theory---not all in complete agreement!--- see D\"{u}rr~ \emph{et al.}~\cite{durr92}, Holland~\cite{holland}, Bohm and Hiley~\cite{bohmandhiley} and Cushing~\cite{cushing94}.) We shall also adopt the convention of referring to the hypothetical particles---the seat of the definite trajectories that are solutions of the guidance equation in the theory---as ``corpuscles'', the reason being that the word ``particle'' has wide usage in all interpretations.} seriously is that, whatever its oddities and weaknesses, it solves the biggest conceptual conundrum in the foundations of the theory: the measurement problem.\footnote{For a recent expression of this attitude, see Myrvold~\cite{myrvold}, pp. 19, 21. See also Cushing's view as discussed in the last section of the present paper.} In the present paper, we wish to take issue with this attitude, or rather its standard construal. 

In our view, de Broglie-Bohm theory does have the resources to provide a coherent solution of the measurement problem, but they do not involve the hypothetical corpuscles whose existence is precisely what distinguishes the theory from the Everettian picture of quantum reality. The standard view is that it is the configuration of the corpuscles in the pilot-wave picture that selects the definite measurement result out of the confusing quantum soup of possibilities that  arises when the initial superposition in the wavefunction description of the object system goes on to infect the state of the entire measuring set-up. We want to argue that this reading of the pilot-wave picture is misleading, and that a proper analysis of the measurement process within the theory should bypass the corpuscles. As a result, the active role of the corpuscles in the theory is called into question. Indeed our view is that if there is a justification for the postulation of the corpuscles, it is not related to the measurement problem \emph{per se}.\footnote{The motivation behind Bohm's original 1952 work is discussed below. In the preface to his 1993 text on de Broglie-Bohm theory, Holland writes
\begin{quotation}\dots our most basic physical theory [quantum mechanics] contains no account of the constitution and structure of matter, corresponding to the interacting particles and fields of classical physics. \dots The aim of the de Broglie-Bohm theory is \textit{not} to attempt a return to classical physics, or even particularly to invent a deterministic theory. \textit{Its goal is a complete description of an individual real situation as it exists independently of acts of observation.}~(\cite{holland}, pp.\,xvii, xviii)\end{quotation}}

This is certainly not the first time the de Broglie-Bohm theory has been adversely compared with the Everett picture; similar conclusions have been reached in writings by Deutsch, Zeh and Wallace.\footnote{See  Deutsch~\cite{deutschlockwood}, Zeh~\cite{zeh} and Wallace \cite{wallacestructure}, section 6. Note that our aim is different from that of Saunders~\cite{saunders99}, who argued that a solution of the measurement problem in relativistic quantum field theory based on a pilot-wave picture in the spirit of de Broglie and Bohm is unsuccessful. Saunders assumed that a solution of the measurement problem based on what we have called the standard construal of the de Broglie-Bohm theory in first-quantised theory is straightforward, and that difficulties only arise when the beables are fields rather than corpuscles. A critique of the standard construal that  is closer in spirit to ours, but based mainly on information-theoretic considerations is found in Stone~\cite{stone}, a hard-hitting reply to which is found in Maudlin~\cite{maudlin}. We return also to the last pair of papers below.} However, we believe these treatments can be strengthened. Before we lay out our case, it is useful to revisit the origins of the theory, or rather David Bohm's early contributions.

\section{Historical considerations}

It is well known that the foundational discussion in David Bohm's text \textit{Quantum Theory}, first published in 1951, is largely consistent with what Bohm would call the ``usual'' interpretation of quantum mechanics---to the extent of containing an argument against the possibility of hidden variable theories.\footnote{Bohm~\cite{bohm51}, section 22.19.} But it is noteworthy that the detailed, conceptually intricate treatment of the measurement process in Chapter 22 is strikingly at odds with the writings of Niels Bohr. Unlike Bohr, Bohm demanded in unequivocal terms a quantum theoretical treatment of the whole  process, including in principle the interaction with the human investigator. (See the quotation at the start of the present paper.) Notable also for its farsightedness is the emphasis Bohm placed on the need for decoherence to occur in the post-measurement state of the joint system comprising object and apparatus systems in order to avoid ``absurd" results.\footnote{\textit{ibid} section 22.11; see also the discussion in section 22.8.}

Bohm's discussion as to why decoherence does in fact occur is not a model of clarity; several distinct arguments seem to be in play.\footnote{See again \emph{ibid} sections 22.8, 22.11} The least satisfactory of these is based on an appeal to a phase-randomization mechanism associated with the measurement process, whose justification and very necessity are both far from obvious. It turns out that Bohm was never entirely happy with this particular argument\footnote{Private communication with Basil Hiley.}. But the point we want to stress is that the reader of his book would be forgiven for concluding that with or without it \emph{Bohm felt he had met the challenge} of providing the bones of an adequate quantum mechanical treatment of the measurement process, without any recourse to a collapse mechanism peculiar to this process. In particular, Bohm did not appear to feel compelled to raise the delicate question as to why the element of the decohered post-interaction wavefunction of the joint system that corresponds to the actual measurement outcome should be favored over the others.\footnote{The closest Bohm comes in his book to addressing the heart of the measurement problem as it is now standardly understood is in section 22.10. Here he argues that despite the joint system being in a pure entangled post-measurement state, for predictive purposes the object subsystem can be described by a "statistical ensemble of separate wavefunctions" (Bohm did not like the word ``mixture"!; p. 604), and since these individual wavefunctions correspond to near-eigenstates of the observable being measured, the process of observation involves nothing more than a gain in information: \begin{quote}When the observer looks at the apparatus, he then discovers in which state the system actually is, by finding out in which of the $\dots$ possible [correlated] classically distinguishable states the observing apparatus is. $\dots$ The sudden replacement of the statistical ensemble of wavefunctions by a single wavefunction represents absolutely no change in the [object system] state $\dots$, but is analogous to the sudden changes in classical probability functions which accompany an improvement in the observer's information. (pp. 603-604)\end{quote} This kind of argument has often been repeated, but it is far from convincing; one familiar way of putting the problem is that Bohm overlooks the distinction between proper and improper mixtures.}

Let's now move on to Bohm's subsequent famous double 1952 paper, in which he independently discovered and developed the essential features of Louis de Broglie's  pilot wave theory of the late twenties. This work patently demonstrated the weakness of Bohm's earlier no-hidden-variables argument, but what if anything was new in the papers in relation to the measurement process?

Much of paper II is concerned with measurement, and builds on the 1951 discussion. It has an explicit treatment---using again a model of impulsive interaction between the microscopic object system and the apparatus---within the Bohm scheme of dynamics for the many-body problem developed in paper I. Bohm states explicitly that the definite coordinate associated with the `apparatus variable' (the hidden variable of the apparatus) will enter one of the non-overlapping wavepackets associated with the final ``$\psi$-field'' of the joint system. The wavepacket in question determines, according to Bohm, the outcome of the measurement. This is a crucial claim---which we shall call the \emph{Result Assumption}--- and we will return to it. We first have to determine what the main point of the discussion of measurement in paper II was.

It is a remarkable and wholly praiseworthy feature of the  ``special assumptions'' defining Bohm's interpretation, and given in section 4 of paper I, that they do not contain the word ``measurement'', and nor does their extension in section 6 to the many-body problem. But eventually measurements must come into the picture and Bohm wrote in the Introduction to paper II: \begin{quote} In this paper, we shall apply the interpretation of the quantum theory suggested in Paper I to the development of a theory of measurements in order to show that as long as one makes the special assumptions indicated above, one is led to the same predictions for all measurements as are obtained from the usual interpretation.\footnote{Bohm~\cite{bohm52}, paper II.}\end{quote}
The aim, then, was with the help also of the Result Assumption, to \emph{match} the ``usual'' interpretation, and not, it would seem, to surpass it. Let us dwell on this point. 

Both 1952 papers make frequent reference to the possibility of altering the ``special assumptions'' so as to produce a theory with truly novel predictions. But the treatment of the measurement process in paper II does not exploit these possibilities, and anyway this is not the feature of Bohm's argument that we wish to underline. Consider rather how Bohm explains that an \emph{effective} collapse of the $\psi$-field (to the packet into which the apparatus coordinate has entered) can be understood from the point of view of post-measurement predictions. Nowhere in this discussion is it explicitly emphasised that since in the theory the true $\psi$-field dynamics of measurement is unitary,  the theory is preferable to any interpretation which contains both unitary and collapse dynamics but fails to explain how they co-exist. Just such a failure is commonly attributed to the Copenhagen interpretation, but as we have seen above in 1951 Bohm appeared to think that  a satisfactory account of measurement was already possible within the usual interpretation, or at least his version of it. At any rate, no hint of superiority is apparent in Bohm's 1952 treatment of the measurement process.\footnote{Of course, Bohm clearly saw his theory as refuting the implication of the usual interpretation ``that we must renounce the possibility of describing an individual system in terms of a single precisely defined conceptual model'' (Paper II, section 10). We return to this point shortly.}  Nor is the above Result Assumption in itself touted as a solution to any sort of problem associated with interpreting the final incoherently superposed $\psi$-field emerging from the measurement interaction. The assumption appears to be simply an element of a new measurement theory the articulation of which is designed simply to \emph{establish predictive compatibility} with the usual interpretation.

Other considerations can be brought to bear on the question as to whether in 1952 Bohm thought that his hidden variable theory solved a conceptual problem related to measurement that the usual interpretation had thrown up.  First, consider Jeffrey Bub's 1997 recollections of his experiences as a graduate student of Bohm in the 1960s. Bub remembers discovering the measurement problem not in discussions with Bohm but through independent reading (particularly of the work of Henry Margenau).\footnote{Bub~\cite{bub}, p. xii.} What surely is even more striking is Bub's recollection that Bohm subsequently suggested to him, in an attempt to solve the measurement problem, an examination of the work of Siegel and Weiner that had appeared in a series of papers in the 1950s. This suggestion was to lead to the 1966 Bohm-Bub hidden variable theory, which explicitly addressed the measurement problem in a fashion that is radically distinct from the ``solution'' that is now widely associated with the pilot-wave theory of 1952. For some years after 1966 Bohm continued to work on a model inspired by the collapse mechanism of the Bohm-Bub theory.\footnote{See \emph{ibid}, p. xii. }

Bohm's own early comments on his 1952 theory, highlighted in Myrvold's recent study\footnote{Myrvold~\cite{myrvold}, \S 3.} of the early response to the theory on the part of the physics community, are also worth mentioning. These comments are noteworthy for their clarification of the motivation behind the theory, and as Myrvold notes, for their emphasis on the provisional nature of his theory. In replying in 1962 to Heisenberg's 1958 critique, Bohm claimed that Heisenberg may have failed to appreciate that 
\begin{quote}$\cdots$the only purpose of this phase of the work was to show that an alternative to the Copenhagen interpretation is at least \emph{logically possible}.\footnote{Bohm~\cite{bohm62a}, p.270.}\end{quote} In the same year he wrote \begin{quote}$\cdots$the existence of even a single consistent theory of this kind showed that whatever arguments one might continue to use against hidden variables, one could no longer use the argument that they are inconceivable.\footnote{Bohm~\cite{bohm62b}, p. 360; \cite{bohm80}, p. 81.} \end{quote} Bohm admitted that his theory ``was not satisfactory for general physical reasons''\footnote{Bohm~\cite{bohm62a}, p.270.}, but saw it as ``a starting point'', the initial step in a process that would eventually lead to a hidden variable theory ``more plausible physically and elegant mathematically''.\footnote{Bohm~\cite{bohm62b}, p. 360; \cite{bohm80}, p. 81.} This diffidence in Bohm's stance to his own 1952 theory is to some extent evident as late as 1993, in his book with Basil Hiley, \emph{The Undivided Universe}\footnote{Bohm and Hiley~\cite{bohmandhiley}, p. 5. Basil Hiley has told us in private communication that Bohm never mentioned his 1952 papers during the first eight or so years of their colloboration in the Department of Physics at Birkbeck College, London. An earlier collaborator of Bohm's at Birkbeck, Chris Philippidis, has likewise told us that he only discovered the existence of the 1952 hidden variable theory while at an overseas conference several years after the start of their colloboration.}. What change there \emph{was} in his thinking between 1952 and 1993 has significantly to do with the theory of measurement. Bohm came to believe that there is a measurement problem in the usual interpretation, probably under the influence of Bub in the sixties, and  from at least 1984 onwards, he and Hiley argued that it finds a solution in the 1952 theory.\footnote{Bohm and Hiley~\cite{bohmandhiley84}; Bohm and Hiley ~\cite{bohmandhiley}, Chapter 6, particularly \S 6.7.}

\section{Bohm's Result Assumption}

It is useful to consider of the precise wording of this assumption. \begin{quote}Now, the packet entered by the apparatus [hidden] variable $\dots$ determines the actual result of the measurement, which the observer will obtain when he looks at the apparatus.\footnote{Bohm~\cite{bohm52}, part II, section 2.}\end{quote}

Suppose we accept that it is the entered wavepacket that determines the outcome of the measurement. Is it trivial that the observer will confirm this result when he or she ``looks at the apparatus''? No, though one reason for the nontriviality of the issue has only become clear relatively recently. The striking discovery in 1992 of the possibility (in principle) of  ``fooling'' a detector in de Broglie-Bohm theory\footnote{See Englert \textit{et al.}~\cite{englert}, Dewdney \textit{et al.}~\cite{dewdney}, Brown \textit{et al.}~\cite{brown95} and Hiley \textit{et al.} ~\cite{hiley00}.} should warn us that it cannot be a mere \emph{definitional} matter within the theory that the perceived measurement result  corresponds to the ``outcome'' selected by the hidden corpuscles. There is much more that needs to be said about this issue, but it is not our central concern.\footnote{We note however that the mentioned possibility of fooling detectors casts doubt on the claim by Maudlin~\cite{maudlin} p. 483, that the so-called effective (post-measurement)  wavefunction of the object system is \emph{defined} (in part) by the positions of the corpuscles associated with the apparatus.}

Our concern rather is with the fact that for Bohm it is the entered \emph{wave packet} that determines the outcome; the role of the hidden variable, or apparatus corpuscle, is merely to pick or select that packet from amongst the other non-overlapping packets in the configuration space associated with the final state of the joint object-apparatus system. How literally Bohm meant this is perhaps open to debate\footnote{Taken literally, Bohm's position in 1952 is not strictly the same as that expressed in Bohm and Hiley~\cite{bohmandhiley}, as we shall see below. It my also be worth pointing out that the notion of corpuscles  `pointing' to or selecting wavepackets is not an entirely sharp one. In the case of a single, more or less localized wavepacket in the configuration space, there is a small chance the corpuscles' configuration is represented by a point well outside the bulk of the packet. It can be said to  `point' to the packet because trivially there is no other, but in the case of a linear combination of near-non-overlapping packets, there must be a very small chance of ambiguous pointing.}, but his wording does invite us to re-examine the apparently innocuous case of \emph{predictable} outcomes, i.e. measurement on a system in an eigenstate of the observable being measured. 

In  this case only a single `localised' wavepacket exists in the configuration space at the end of the measurement process---the  wavepacket correlated with the initial eigenvector of the observable being measured which happens to be the initial state of the object system. The crucial question we wish to raise is this. \emph{Does  this wavepacket, in and of itself, account for the result of the measurement, or does a definite measurement outcome require, even in this case of complete predictability, the presence of the hidden variables within it?}

Most discussions of the measurement problem in quantum mechanics take it for granted that no difficulties arise in this case of the predictable outcome---that the problem only rears its head in the more interesting and more general case of unpredictability, when the intial state of the object system is some linear combination of eigenvectors of the relevant observable. But if analysis of the predictable case is successful without appeal to hidden variables, then Bohm's Result Assumption in the general case is problematic. In the general case,  \emph{each of the non-overlapping packets in the final joint-system configuration space wavefunction has the same credentials for representing a definite measurement outcome as the single packet does in the predictable case.} The problem, if it is one, is that there is more than one of them. But the fact that only one of them carries the de Broglie-Bohm corpuscles does nothing to remove these credentials from the others. Adding the corpuscles to the picture does not interfere destructively with the empty packets. The Result Assumption appears to be inconsistent with the treatment of the predictable case, or at least to override it in some mysterious way.

It seems to us that the only way to avoid this conundrum in the de Broglie-Bohm theory is to adopt the position that in the predictable case deriving the single localised wavepacket is \emph{not} sufficient to account for the outcome of the measurement. Just such a position has been defended by Holland. In his monumental text \emph{The Quantum Theory of Motion}, he writes in relation to the measurement problems and the Schr\"{o}dinger cat paradox that even if some stochastic, non-unitary collapse occurs \begin{quote} $\cdots$ the definite state so obtained will still only be a wavefunction that in itself exhibits no feature that may be identified with the reality of a definite pointer position, or the centre of mass of a cat.\footnote{Holland~\cite{holland}, p. 334.} \end{quote} This point is reinforced in Holland's detailed discussion of the dynamics of the measurement process, where he distinguishes two aspects of the coupling of the object system and apparatus: what it takes to get a definite outcome and what it takes to ensure that this outcome carries information about the object system. \begin{quote} The definiteness of the final outcome is a property of the definiteness of the [apparatus] pointer location [defined by the configuration of apparatus corpuscles] under all circumstances (i.e. whatever the quantum state). This is the crucial point, and not that $\Psi$ is composed of a set of disjoint packets. The latter is simply a necessary condition that allows us to ascertain from the always well-defined pointer reading unambiguous information on an object property, and is not itself the condition for definiteness.\footnote{\emph{ibid}, p. 347} \end{quote} Concluding his analysis of the measurement process, Holland writes \begin {quote} It is the assumption of a corpuscle which transforms quantum mechanics into a theory having substance and form. The pure wave dynamics described by Schr\"{o}dinger's equation does not yield any account of which result is actually realized in an individual measurement operation. The wavefunction collapse hypothesis only gains physical content if actual coordinates for the collapsed system are posited. Since the point at which these are introduced in the chain of connected physical systems is arbitrary, the only consistent assumption is that they are well defined all along.\footnote{\emph{ibid}, p. 350} \end{quote}
So it is only  by adopting Holland's position that the Result Assumption makes sense in the de Broglie-Bohm theory. However, we resist Holland's position. Quantum mechanics, in our view, has both substance and form before the introduction of hidden corpuscles. In fact, we believe the case has already been made elsewhere, the rest of this paper being an attempt to summarise the bones of the argument and to add a bit more new flesh to it. To this end, the case of the single-outcome, predictable measurement will be repeatedly exploited.

\section{The main argument}

Comparison of the de Broglie-Bohm theory with the Everett interpretation will be the starting point of our argument. For from the perspective of mathematical physics it seems (at
least at first sight) to be rather odd to claim that the former is ontologically
more frugal than the latter. The mathematical formalism of the Bohm theory, after all,
consists of two elements: the quantum state vector, or wave function, evolving according to the
Schr\"{o}dinger equation, and the corpuscles, represented by a point in configuration
space. The Everett interpretation requires only the first of these two elements ---
so why is it not Bohm's theory, rather than Everett's, which contains excess ontology?

Until perhaps the late 1980s, Bohmians had a ready answer to this question. The ``many worlds''
talk used in discussions of Everett (they might say) is \emph{mere} talk unless the formalism 
is supplemented by some additional mathematical structure. If such supplementation
is attempted (for instance, by Deutsch's~\cite{deutsch85} explicit specification 
of the ``worlds'', or by Albert and Loewer's \cite{albertloewermm}
addition of ``minds'' whose trajectories were determined by the state
vector\footnote{For criticisms of these two views, see respectively Foster and Brown~\cite{foster} and Lockwood~\cite{lockwood}.}) then the ontological comparison is indeed in Bohm's favour: in
place of a single $N$-tuple of corpuscles the Everett interpretation has a
vast collection of such $N$-tuples, representing countless worlds, or
countless minds. Why not then abandon all but one, and trim a bloated
ontology?

However, in recent years Everett's defenders\footnote{See, e.g.\,, Saunders \cite{saundersdecoherence}, Wallace \cite{wallaceworlds,wallacestructure}, or Vaidman \cite{vaidman}.} have been less willing to
concede the necessity of supplementing the wave-function with any
additional structure. Instead, they have increasingly defended the view
that the ``worlds'' are in some sense emergent from the bare 
wave-function, in rather the same way that everyday objects in our
macroscopic world (cats, say; or tables) emerge from the microphysics
without having to be explicitly added to the formalism. The main
technical development behind this shift has been decoherence theory,
which provides a robustly-defined (if not quite perfectly defined)
choice of preferred basis with which to defined the worlds. ``Worlds''
in this sense can be seen as decoherence-defined components of the
wave-function, effectively though not quite perfectly immune from
interference with other such components.

From this modern perspective, the ontological situation shifts. The
Everettian ontology is now robustly monistic --- the wave-function
constitutes the whole of reality --- and it is the Bohm theory which 
is ontologically excessive.

This criticism is only sharpened by the recognition that decoherence 
--- central to the modern view of Everett --- is also essential in the
Bohmian picture. Although the Bohmian corpuscle picks out by fiat a
preferred basis (position), the de Broglie-Bohm theory still has to tell some story
about the measurement-induced effective collapse of the wave function. Bohm
recognised this in 1952; from a
modern perspective, we recognise this as the requirement that (1)
decoherence occurs, and (2) the preferred basis which it picks out is
(approximately) the position basis.

From this viewpoint, the corpuscle's role is minimal indeed: it is in
danger of being relegated to the role of a mere epiphenomenal `pointer', 
irrelevantly picking out one of the many branches defined by decoherence, 
while the real story --- dynamically and ontologically --- is being told
by the unfolding evolution of those branches. The ``empty wave packets'' in the configuration space
which the corpuscles do not point at are none the worse for its absence:
they still contain cells, dust motes, cats, people, wars and the like.
The point has been stated clearly by Zeh:\begin{quote} It is usually overlooked that Bohm's theory contains the \emph{same} ``many worlds'' of dynamically separate branches as the Everett interpretation (now regarded as ``empty'' wave components), since it is based on precisely the same $\dots$ global wave function $\dots$\footnote{Zeh~\cite{zeh}. In this paper, Zeh states on the basis of private correspondence with him in 1981 that ``John Bell (who rejected Everett's interpretation for being ``extravagant''), seems to have realized this equivalence before he began to favor spontaneous localization (such as GRW) over Bohm's theory $\dots$".}\end{quote} Deutsch has expressed the point more acerbically:
\begin{quote}
[P]ilot-wave theories are parallel-universe theories in a state of
chronic denial.\footnote{Deutsch~\cite{deutschlockwood}.}
\end{quote}

\section{Decoherence}

How can the de Broglie-Bohm theory be defended against this sort of attack? In this
section and the next two, we will explore what we see as the range of defences
available and argue that none succeeds.

Perhaps the most obvious --- and the most principled --- defence is to attack
the coherence of the Everettian account on its own terms: if
(notwithstanding decoherence) we cannot regard worlds as emergent from
the wavefunction, then the corpuscle retains its role as the definer of
those worlds and the wavefunction remains simply an auxiliary field ---
albeit an astonishingly complicated one.

Such an attack might be mounted in one of three ways. The first is
purely technical: it might be that decoherence does not in fact succeed
in producing a preferred basis (and specifically a quasi-classical
basis) in realistic models. This is a long shot, however, given the
impressive success of the decoherence program for simpler (but
increasingly complicated) models; as mentioned above, it also threatens
to undermine the Bohm theory, which is also highly dependent on
decoherence.\footnote{See Bohm and Hiley~\cite{bohmandhiley}, section 6.1.}

The second and third attacks are more philosophical in nature. The
second is as follows: granted that decoherence picks out a 
quasi-classical basis as preferred, what is to say that it does not also
pick out a multitude of other bases --- very alien with respect to the
bases with which we ordinarily work, perhaps, but just as `preferred'
from the decoherence viewpoint. Such a discovery would seem to undermine
the objectivity of Everettian branching, leaving room for the Bohmian
corpuscle to restore that objectivity.

Saunders~\cite{saundersdecoherence} offers an anthropic response to this attack.
Suppose that there were several such decompositions, each supporting
information-processing systems. Then the fact that we observe one rather
than another is a fact of purely local significance: we happen to be
information-processing systems in one set of decoherent histories rather
than another.

There is perhaps a more robust response
available to the Everettian here. Granted that we cannot rule out the possibility 
that there \emph{might be} alternative decompositions, and that this
would radically affect the viability of the Everett interpretation 
--- well, right now we have no reason at all to suppose
that there actually \emph{are} such decompositions. Analogously, 
logically we can't absolutely rule out the possibility that there's a
completely different way of construing the meaning of all English words,
such that they mostly mean completely different things but such that
speakers of English still (mostly) make true and relevant utterances.
Such a discovery would radically transform linguistics and philosophy,
but we don't have any reason to think it will actually happen, and we
have much reason to suppose that it will not. To discover one sort of
higher-level structure in microphysics (be it the microphysics of
sound-waves or the micro-physics of the wave-function) is pretty
remarkable; to discover several incompatible structures in the same bit
of microphysics would verge on the miraculous.\footnote{See Dennett~\cite{dennett93}, pp. 344--348, for more on these ``cryptographic constraints'' on the indeterminacy of
translation.}

The third attack involves the approximate nature of the 
decoherence-preferred basis. Reality (it says) should not be an
approximate, fuzzy thing: a criterion for macroscopic definiteness
should be sharp and well-defined. This has been popular with a number of
physicists, including Bell~\cite{bell}, Kent~ \cite{kent} and as we have seen, Holland.

However, there seems no obvious basis for such a requirement.\footnote{See Wallace \cite{wallacestructure}.} In
general, macroscopic objects are not precisely defined in microphysical
terms: rather, they emerge from the microphysics as robust structures or
patterns as, for example, Dennett~ \cite{realpatterns} has argued --- thus, a cat is a higher-level
structure in a lower-level ontology of cells, and those cells are in
turn structures in a lower-level ontology of atoms and molecules. It is
in the nature of such structures to be `noisy' --- that is, very robust
but not perfectly so. This view of higher-level ontology has been called a
`functionalist view' by analogy with (and as a generalisation of) the 
related view in philosophy of mind to which we return in section 7. It is fair to point out that it has
come under some criticism in recent years from metaphysicians\footnote{In
particular Jaegwon Kim; see for instance ~\cite{kim98}.}, but it is also 
fair to describe it as virtually a consensus among philosophers, and
philosophically inclined workers, in cognitive science and the
philosophy of the special sciences.\footnote{For a recent review of the debate,
see Ross and Spurrett~\cite{rossspurrett}, and the comments on their paper in a forthcoming
issue of \emph{Behavioral and Brain Sciences}.}

For these reasons we find the attacks on the \emph{coherence} of the
Everettian ontology unpersuasive, committed as they are either to
conflict with mainstream views in physics and in general philosophy.

\section{Degrees of Reality?}

If Bohmians cannot deny that the `empty waves' contain real macroscopic
objects on grounds of \emph{structure}, they may be able to do so from a more metaphysical
perspective. There are two possible ways that they might do so: firstly, they
might deny that the wave-function is the `sort of stuff' from which cats, tables
and physicists can be made; alternatively, they might take the bolder step of denying
the reality of the wave-function altogether.

Something like the first position is defended in, say, Bohm and Hiley's  book \emph{The Undivided Universe}: \begin{quotation} $\dots$ it could be felt that the empty packets, which also satisfy Schr\"{o}dinger's equation, constitute a vast mass of `bits of reality' that are, as it were, `floating around' interpenetrating that part of reality which corresponds to the occupied packets.

In our interpretation, however, we do not assume that the basic reality is thus described primarily by the wavefunction. Rather $\dots$ we begin with the assumption that there are \emph{particles} following definite trajectories. $\dots$ We then assume that the wavefunction, $\psi$, describes a qualitatively new kind of quantum field which determines the guidance conditions and the quantum potential acting on the particle. We are not denying the reality of this field, but we are saying that its significance is relatively subtle in the sense that it contains active information that 'guides' the particle in its self-movement under its own energy. $\dots$ So ultimately all manifestations of the quantum fields are through the particles.\footnote{Bohm and Hiley~\cite{bohmandhiley}, pp. 104-5.} \end{quotation} 

The electromagnetic field is also only detectable by way of its action on `matter'. (Indeed, since Bell, it has been popular to regard the wave function in the Bohm theory as a physical field akin to the electromagnetic field.) But the reason we regard the electromagnetic field as immaterial, in so far as we do, is that \emph{being linear} it is not rich enough alone to support the sorts of complex structures that
give rise to macroscopic objects. Not long ago it was fashionable to
speculate about large-scale structures being built out of (nonlinear) pure geometry;
it is still fashionable to speculate that nonlinearities in quantum
chromodynamics could give rise to objects built purely out of the gluon
field. If we encountered objects like these, we would not deem them
``immaterial'' simply because they were made out of the wrong sort of
field.\footnote{In any case, in quantum field theory the matter/field
distinction breaks down to a substantial degree. The observed mass of
any `material' particle owes as much to its cloud of virtual field
particles as it does to its material components.} And let us not forget that the pure electromagnetic field is capable on its own of bending space-time---and that there is every reason to think that the wavefunction is too.

We are hard-pressed then to see why the wavefunction should be regarded as a second-class citizen in the reality stakes. A slightly different tack is taken by Bohm and Hiley in their discussion of the Schr\"{o}dinger cat paradox, which attempts to address the ``ambiguity of the state of being" of the cat when it finds itself entangled with the killing machine, the joint system having been brought in the familiar way into a superposition of the dead and alive possibilities. \begin{quotation} To define the actual state of being of the cat, we have to consider in addition the particles that constitute it. It is evident that when the cat is alive, many of these particles will be in quite different places and will move quite differently than they would if the cat were dead.

$\dots$ in our interpretation that state of being also depends on the positions of the particles that constitute [the cat] $\dots$ and it is this which enables us to treat this situation non-paradoxically as in essence a simple case of pairs of alternative and mutually exclusive states.\footnote{ibid. pp. 126--7.}\end{quotation}
We cannot but agree that the cat-corpuscles will find themselves in one and only one element of the superposition, and that their configuration and motion will depend on which, but does this mean that the ``state of being'' of the cat is singular in the sense Bohm and Hiley are after?

Consider the element in the superposition that does not contain the corpuscles---the `empty' wavepacket in the configuration space of the joint system. Does it not describe (amongst other things) a cat that is either dead or alive? Yes, most of us would say, if the cat had actually been prepared exclusively in the cat-state that is contained in (better: is a factor state of) that element. Recalling the discussion in section 3 above of the predictable single-outcome measurement, and the arguments of the previous section, is it not more natural to say that the superposition describes both a dead and an alive cat (each correlated with distinct states of the environment) with one of these possibilities replete with corpuscles? It is hard to see how the corpuscles annul the reality of the other possibility; indeed they cannot. To argue that the ``state of being'' of the cat depends ``in addition" or ``also" on the corpuscles is to admit that the wavefunction plays some role in the matter, but not one consistent with the common interpretation of the predictable, single outcome measurement scenario.  

It is interesting that at times Bohmians slide into a mode of talking about systems as if they were \emph{just} made out of corpuscles\footnote{See for example, Bohm and Hiley~\cite{bohmandhiley} in the final paragraph of section 6.7, or the quotation from Maudlin~\cite{maudlin} given in the next section of the present paper.}, but this is only coherent if the radical position is adopted that the wavefunction is simply not real at all, that it is a piece of mathematical machinery in the quantum mechanical algorithm for the motion of corpuscles. Yet `reality' is not some property which we can grant or 
withhold in an arbitrary way from the components of our mathematical
formalism. The wave-function evolves; it dynamically influences the
corpuscles; in interference experiments its existence is explanatorily
central to the observed phenomena. On what grounds could we just dismiss
it as a mathematical fiction?

Again there is a bad analogy to resist here. From the corpuscles'
perspective, the wave-function is just a (time-dependent) function on
their configuration space, telling them how to behave; it superficially
appears similar to the Newtonian or Coulomb potential field, which is again a
function on configuration space. No-one was tempted to reify the
Newtonian potential; why, then, reify the wave-function?

Because the wave-function is a very different sort of entity. It is
contingent (equivalently, it has dynamical degrees of freedom
independent of the corpuscles); it evolves over time; it is structurally
overwhelmingly more complex (the Newtonian potential can be written in
closed form in a line; there is not the slightest possibility of writing 
a closed form for the wave-function of the Universe.) Historically, it
was exactly when the gravitational and electric fields began to be
attributed independent dynamics and degrees of freedom that they were
reified: the Coulomb or Newtonian `fields' may be convenient
mathematical fictions, but the Maxwell field and the dynamical spacetime
metric are almost universally accepted as part of the ontology of modern
physics. 

We don't pretend to offer a systematic theory of which mathematical entities 
in physical theories should be reified. But we do claim that the decision is not
to be made by fiat, and that some combination of contingency, complexity and time
evolution seems to be a requirement.\footnote{Those who sympathise with Leibniz' claim---fully endorsed by Einstein---that the essence of a real thing is its ability to act and be acted upon, may be interested in a defence of the reality of the wavefunction based on the action-reaction principle found in Anandan and Brown~\cite{anandan}.} 

To illustrate further why denying reality to the wave-function by fiat
would go against basic principles of physics, consider Wheeler and
Feynman's heroic attempt to eliminate the electromagnetic field. It is
generally recognised that they failed (not least because of the advent
of quantum electrodynamics) but that they approached the problem in an
appropriate way: find a way of rewriting the dynamical equations such
that the electromagnetic field entirely disappears, and make whatever
hypotheses about the distant past we require in order to justify the
rewrite. How much easier --- but how much more unprincipled --- it would
have been for them simply to declare that the electromagnetic field does not
exist, but that matter behaves \emph{as if} it did exist.

This analogy does suggest one possible strategy for Bohmians, however:
rather than just denying the reality of the wave-function, find a
principled way to eliminate it from the theory and recover it as an
effective, phenomenological object (just as Wheeler and Feynman
attempted to do for the electromagnetic field). Such a strategy has been 
advocated by Goldstein and co-workers~\cite{goldstein,goldsteinteufel}, taking its motivation from
quantum cosmology. They suppose that the wave-function of the Universe
(regarded as the solution of some Wheeler-de Witt-type equation) may
turn out to be both unique and time-independent, and that this will make
it appropriate to regard it more as a physical law than as a physical
object. 

Two comments should be made about this research program.

Firstly, it \emph{is} a research programme. Bohmian
quantum cosmology, as with all other proposals for quantum cosmology, is
at present purely speculative, beset with technical and conceptual
problems, and quite disconnected from experiment. (Golstein and Teufel~\cite{goldsteinteufel} give a sympathetic review of its current
status.) As such, the existence \emph{right now} of a Bohmian solution of the measurement problem based on ideas from quantum cosmology is as implausible as the notion that Penrose's speculative suggestions
about gravitation-induced collapse mean that the measurement problem is now 
solved.

Secondly, and perhaps more importantly, even \emph{given} a technically satisfactory Bohmian cosmology, 
it is by no means clear that we should not reify that cosmology's wavefunction.
We identified earlier three features of the wavefunction which 
distinguish it from the Newtonian potential: it is dynamical; it is
contingent; and it is extremely richly structured. Of these, it is at best very unclear
that any of them fail in a cosmological context. As far as dynamics are concerned, though
the cosmological wavefunction is usually taken to be time-independent, the notion
of time is so controversial in quantum cosmology that we would be
reluctant to jump too quickly from this time-independence to any claim about dynamics.
As for contingency, it is at most an article of faith
with some physicists that the Wheeler-de Witt equation has a unique
solution.
And as for complexity (in our view perhaps the most important criterion)
the structure encoded in the cosmological wavefunction will
if anything be richer than that encoded in the nonrelativistic
wavefunction.

\section{Consciousness}

If it correct that the the de Broglie-Bohm theory essentially involves a dualist ontology, with corpuscle \emph{and} wavefunction on an equal footing, there is nonetheless a fundamental asymmetry in the duality. As Bell stressed, ``It is $\dots$ from the \textbf{x}s [the coordinates of the corpuscles], rather than the $\psi$, that in this theory we suppose `observables' to be constructed''.\footnote{Bell~\cite{bell}, p. 128.} Could this asymmetry be the basis of another strategy for the Bohmians?

Observables in the context of Bell's remark are defined relative to sentient observers, and it is a tenet of the de Broglie-Bohm picture that such observers are aware of corpuscles in a way that fails to hold for wave functions. Of course, there is an obvious sense in which the corpuscles are also `hidden', and D\"{u}rr \emph{et al.} emphasized in 1992 that the only time we can have sure knowledge of the configuration of corpuscles is ``when we ourselves are part of the system''.\footnote{D\"{u}rr \emph{et al.}~\cite{durr92}.} But how exactly is this supposed to work? Stone correctly pointed out in 1994 that this claim certainly fails if our knowledge is based on measurements which one part of our brain makes on another.\footnote{Stone~\cite{stone}.} But perhaps what D\"{u}rr \emph{et al.} had in mind was closer to Maudlin's 1995 position. \begin{quote} If we want to know what happened to the measuring device (e.g., which way the pointer went), we look at it, thereby correlating positions of particles in our brains with the pointer position. If getting the state of our brain correlated with the previously unknown external conditions is not getting information about the world, nothing is.\footnote{Maudlin~\cite{maudlin}, p. 483.}\end{quote} This last sentence is unimpeachable, and it applies as much to the Everett as to the de Broglie-Bohm interpretation.  But Maudlin seems to be taking it for granted that our conscious perceptions supervene directly \emph{and exclusively} on the configuration of (some subset) of the corpuscles associated with our brain. If this is so, it is true we gain information by way of such correlations between corpuscles in and outside of our craniums. But the question is, why believe this theory of psycho-physical parallelism? 

Recall Bohm's 1951 exortation to allow inclusion of the human investigator into the quantum-mechanically treatable measurement chain, and consider once again its significance in the case of the \emph{predictable} measurement raised above in section 3. Few  would argue that in the absence of de Broglie-Bohm corpuscles quantum mechanics is incapable in principle of accounting for the predictable sensation on the part of the sentient observer. Why shouldn't consciousness supervene as much on wavefunctions as on corpuscles?---a possibility that was clearly entertained by Bohm in 1951. But if one allows for this possibility, the floodgates into the Everettian multiplicity of autonomous, definite perceived outcomes are opened.\footnote{As we saw in section 3, even in his hidden variables paper II of 1952, Bohm seems to associate the wavepacket chosen by the corpuscles as the representing outcome of the measurement---the role of the corpuscles merely being to point to it. But if this wavepacket can support consciousness, it is mysterious why empty ones cannot.} To restrict supervenience of consciousness to de Broglie-Bohm corpuscles in the brain does succeed in restricting conscious goings-on to one and only one branch of the Everett multiverse but it seems unwarranted and bizarre. The strategy seems unmotivated except by a desire purely
to reduce the number of conscious observers in the universe, and it is at best unclear
whether this is a reasonable application of Occam's Razor.\footnote{It is noteworthy that the \emph{active} role of the corpuscles in the de Broglie-Bohm theory is merely to act on each other, not back on the wavefunction. So it is striking that such passive entities are purportedly capable of grounding consciousness experience. \emph{C.f.} Stone~\cite{stone} and independently Brown~\cite{brown96}.} We do not in
general choose between competing evolutionary theories on the grounds of
minimising the number of predicted alien civilisations, nor between competing 
theories of human prehistory on the grounds of minimising the number of
\emph{homo sapiens} who have walked the earth. And the price that is paid is the existence of a vast number of zombies or mindless hulks (you choose your nomenclature) associated with the empty waves that act precisely as if they were conscious but are not.

It is worth bearing in mind too that significant coarse-graining must take place in the process of our awareness of the corpuscles. Considerable variation in the precise configuration of the corpuscles in our brain must be unobservable. How much? Not even the effective wave function of our brain can be directly known precisely; even Everettians must admit there is a many-one relationship between wave functions of the brain and conscious sensations. But it seems to be the case that the effective wave function of the brain represents a strict \emph{lower bound} on the course-graining of consciousness. A violation of the no-signalling theorem is possible in principle were we to `know' the configuration of corpuscles in our brain with a greater level of accuracy than that defined by the wave function. To our knowledge this important point was first made by Valentini in 1992.\footnote{Valentini~\cite{valentini}, p. 30.} He wrote further:\begin{quote} This might seem to argue in favour of abandoning the hidder-variable altogether, and retaining only the wavefunction, thereby leading to the many-worlds theory. But the situation is really the same here as in classical mechanics: An experimenter built from classical atoms has a reality which rests on the precise configuration $\phi$ of all his atoms, and yet his functioning is completely insensitive to $\phi$, beyond a certain level of accuracy $\dots$\footnote{\emph{ibid}, pp. 30--31.}\end{quote}.
It is not clear to us that the analogy with classical mechanics is accurate, though. An
experimenter ``built from classical atoms" has an approximate configuration which is \emph
{logically supervenient} on his exact configuration: it doesn't make sense to suppose one specified
without the other. By contrast, in the de Broglie-Bohm theory the ``approximate configuration" defined by the wave-packet is a completely distinct entity from the ``exact configuration" defined by the corpuscles, and we can imagine either existing independently of the other.

ÊIn any case, Valentini's point reminds us of an important---but extremely questionable---assumption about the nature of consciousnes which our discussion thus far has been making: namely, the assumption that consciousness is some sort of bare physical property (like, say, charge) whose connection with physical matter can simply be posited rather as we posit other basic physical laws. To be sure, this approach to consciousness is not uncommon in the foundations of physics: arguably it was first made explicit in von Neumann's doctrine of psycho-physical parallelism, which suggests that the link between conscious experience and the physical world is just one more thing to be specified by any interpretation of quantum mechanics.

ÊSuch an approach, however, makes consciousness completely divorced from any assumptions rooted in the study of the brain. In neuroscience, in cognitive science, in AI, and in psychology, it is almost universally assumed that consciousness is some \emph{emergent} property of physical matter, the emergence of which is to be understood ``functionally'' --- that is, in terms of how that physical matter moves and interacts. If this were not the case, then --- given the causal closure of the physical world --- consciousness could have no empirical consequences, and would be opaque to third-person scientific study.\footnote{See \cite{dennettzombic} for further development of this point.} If on the other hand the functionalist assumption is correct, for consciousness to supervene on the corpuscles but not the wavepackets, the corpuscles must  have some functional property that the wave packets don't share. But as we have just seen, the functional behaviour of the corpuscles is identical to that of the wavepacket in which they reside.

This is not to say that Bohmians who want to reject the functionalist assumption for consciousness wouldÊ be entirely without allies. Among scientists, while neuroscientists, psychologists \emph{et al.} are hostile to the approach, a few physicists have been substantially more sympathetic.\footnote{One commentator who effectively rejects the functionalist approach to consciousness, by claiming that the Everett picture is simply opaque as to what perceptions should be like, is Penrose; see his \cite{penrose89} p. 296 and \cite{penrose94} p. 312. How, Penrose asks, do we rule out the emergence of a conscious view of the world that corresponds to bases other than the decoherence basis and relative to which it seems that all hell breaks loose, or at least macroscopic things like elephants seem to smear out over space? Yet we note that  the principle of democracy of orthogonal bases applies just as much to the single-outcome predictable case. When we can predict with certainty that an observer will be aware of a certain outcome, it seems we do not get into a sweat over the fact that the usual wavefunction describing the predicted state of affairs (including the mental state of the observer) could also be written as a linear combination of bizarre-looking states belonging to an arbitrary basis in the Hilbert space associated with the joint system. It is noteworthy that Penrose uses precisely the single-outcome measurement to argue for the reality of the wave function (\cite{penrose94}, p. 315.), and it would seem odd were he not to see the relation in this case between the state vector and what we are supposed actually to observe. (Certainly Penrose, unlike Holland, does not mention any need to introduce further structure into quantum mechanics to this end; indeed to do so would be incompatible with the spirit of Penrose's own gravitationally-induced collapse model.) But if this relation is transparent in the predictable case, as it seems to be, then its alleged opaqueness in the general case is not a foregone conclusion. To repeat the point made in section 3, each element of the relevant decoherence basis has the same credentials to represent a definite, familiar conscious state on the part of the observer as the single element does in the predictable case.} Among philosophers, some eminent voices (such as Searle \cite{searlerediscovery} and Chalmers \cite{chalmers}, and in the context of quantum mechanics, Lockwood \cite{lockwood}) have rejected the functionalist approach and held out for an alternative, first-person-centred, science of the mind. But it should be recognised that these are distinctly unorthodox opinions in both cases.

Ê
\section{Conclusions}

James Cushing was an eloquent and indefatigable promoter of the de Broglie-Bohm theory. In an article he published in 1996 entitled ``What Measurement Problem?'', Cushing wrote the following.
\begin{quote} One of the most beautiful aspects of Bohm's [\cite{bohm52}, Paper II] $\dots$ is his treatment of the measurement problem (which becomes a non-problem). In his theory, measurement is a dynamical and essentially a many-body process. There is \emph{no} collapse of the wavefunction and, hence, no measurement problem. The basic idea is that a particle always has a definite position between measurements. There is no superposition of properties and ``measurement'' (or observation) is an attempt to discover this position.\footnote{Cushing~\cite{cushing96}, p. 169. Despite its title, this paper is mostly concerned with the issue of compatibility or otherwise of the de Broglie-Bohm theory with relativity theory.} \end{quote}

We agree that Bohm's 1952 treatment of the measurement process is beautiful, as is his 1951 treatment. Both effectively attempt to show that not every interpretation of quantum mechanics needs to introduce the term ``measurement'' as a primitive.  Although we are doubtful as to whether Bohm in 1952 intentionally addressed the ``measurement problem'' in its standard sense, we agree with the claim that since there is no collapse of the wavefunction in the de Broglie-Bohm theory, there is no measurement problem. But we disagree with the final two sentences in this quote.

For in these sentences, Cushing enshrines the hypothetical corpuscle (or its ``definite position'') as the foundation of the ``basic idea'' of measurement. In our view, the existence of such an entity does \emph{not} mean there is no superposition of properties. This superposition is a fact, and Everett's greatest legacy is to have suggested that appearances can be saved despite it.  Indeed we have tried to argue that observation---in so far as this is related to the cognitive process of  ``knowing'' the outcome of the measurement process---is \emph{not} discovering the position of the de Broglie-Bohm corpuscle even if it exists.

\section{Acknowledgements}

It was unfortunately the lot of only one of us (HRB) to know James Cushing, and he is grateful to the Editors to have this opportunity to express his admiration of both the scholar and the man. 

We are also grateful to Jeffrey Bub, Basil Hiley, Peter Holland, Chris Philippidis and Antony Valentini for extremely useful discussions and/or correspondence.

\end{document}